\documentclass[aps,prl,twocolumn,unsortedaddress,superscriptaddress,nobibnotes]{revtex4}
\usepackage{color,graphicx}
\usepackage{array,amsmath,amssymb,pifont,mathrsfs,latexsym}

{\begin{list}{}{%
      \settowidth{\labelwidth}{\textbf{#1}}%
      \setlength{\leftmargin}{\labelwidth}
      \addtolength{\leftmargin}{0.5em}}}
  {\end{list}}
\newcommand{\nuc}[2]{\ensuremath{^{\text{#1}}\text{#2}}}

\newcommand{\CaFCaF}{\nuc{40}{Ca}+\nuc{40}{Ca}}
\newcommand{\CaFECaFE}{\nuc{48}{Ca}+\nuc{48}{Ca}}
\newcommand{\SnSnT}{\nuc{112}{Sn}+\nuc{112}{Sn}}
\newcommand{\SnSnF}{\nuc{124}{Sn}+\nuc{124}{Sn}}

\newcommand{\degree}{\ensuremath{^\circ}}

\begin{document}

\title{Scaling properties of light-cluster production}
\author{Z. Chaj\c{e}cki}
\email[Email comments to: ]{chajecki@nscl.msu.edu}
\affiliation{National Superconducting Cyclotron Laboratory, Michigan State University, East Lansing, MI 48864, USA}
\author{M. Youngs}
\affiliation{National Superconducting Cyclotron Laboratory, Michigan State University, East Lansing, MI 48864, USA}
\affiliation{Department of Physics and Astronomy, Michigan State University, East Lansing, MI 48864, USA}
\author{D.D.S. Coupland}
\affiliation{National Superconducting Cyclotron Laboratory, Michigan State University, East Lansing, MI 48864, USA}
\affiliation{Department of Physics and Astronomy, Michigan State University, East Lansing, MI 48864, USA}
\author{W.G. Lynch}
\affiliation{National Superconducting Cyclotron Laboratory, Michigan State University, East Lansing, MI 48864, USA}
\affiliation{Department of Physics and Astronomy, Michigan State University, East Lansing, MI 48864, USA}
\affiliation{Joint Institute of Nuclear Astrophysics, Michigan State University, East Lansing, MI 48864, USA}
\author{M.B. Tsang}
\affiliation{National Superconducting Cyclotron Laboratory, Michigan State University, East Lansing, MI 48864, USA}
\affiliation{Department of Physics and Astronomy, Michigan State University, East Lansing, MI 48864, USA}
\affiliation{Joint Institute of Nuclear Astrophysics, Michigan State University, East Lansing, MI 48864, USA}
\author{D. Brown}
\affiliation{National Superconducting Cyclotron Laboratory, Michigan State University, East Lansing, MI 48864, USA}
\affiliation{Department of Physics and Astronomy, Michigan State University, East Lansing, MI 48864, USA}
\author{A. Chbihi}
\affiliation{GANIL, CEA et IN2P3/CNRS, F-14076 Caen, France}
\author{P. Danielewicz}
\affiliation{National Superconducting Cyclotron Laboratory, Michigan State University, East Lansing, MI 48864, USA}
\affiliation{Department of Physics and Astronomy, Michigan State University, East Lansing, MI 48864, USA}
\affiliation{Joint Institute of Nuclear Astrophysics, Michigan State University, East Lansing, MI 48864, USA}
\author{R.T. deSouza}
\affiliation{Department of Chemistry, Indiana University, Bloomington, IN 47405, USA}
\author{M.A. Famiano}
\affiliation{Department of Physics, Western Michigan University, Kalamazoo, MI 49008, USA}
\author{T.K. Ghosh}
\affiliation{Variable Energy Cyclotron Centre, 1/AF Bidhannagar, Kolkata 700064, India}
\author{B. Giacherio}
\affiliation{Department of Physics, Western Michigan University, Kalamazoo, MI 49008, USA}
\author{V. Henzl}
\affiliation{National Superconducting Cyclotron Laboratory, Michigan State University, East Lansing, MI 48864, USA}
\author{D. Henzlova}
\affiliation{National Superconducting Cyclotron Laboratory, Michigan State University, East Lansing, MI 48864, USA}
\author{C. Herlitzius}
\affiliation{National Superconducting Cyclotron Laboratory, Michigan State University, East Lansing, MI 48864, USA}
\affiliation{Joint Institute of Nuclear Astrophysics, Michigan State University, East Lansing, MI 48864, USA}
\author{S. Hudan}
\affiliation{Department of Chemistry, Indiana University, Bloomington, IN 47405, USA}
\author{M. A. Kilburn}
\affiliation{National Superconducting Cyclotron Laboratory, Michigan State University, East Lansing, MI 48864, USA}
\affiliation{Department of Physics and Astronomy, Michigan State University, East Lansing, MI 48864, USA}
\author{Jenny Lee}
\affiliation{National Superconducting Cyclotron Laboratory, Michigan State University, East Lansing, MI 48864, USA}
\affiliation{Department of Physics and Astronomy, Michigan State University, East Lansing, MI 48864, USA}
\author{F. Lu}
\affiliation{Joint Institute of Nuclear Astrophysics, Michigan State University, East Lansing, MI 48864, USA}
\affiliation{Shanghai Institute of Applied Physics, Chinese Academy of Sciences, Shanghai 201800, China}
\author{S. Lukyanov}
\affiliation{National Superconducting Cyclotron Laboratory, Michigan State University, East Lansing, MI 48864, USA}
\affiliation{FLNR, JINR, 141980 Dubna, Moscow region, Russian Federation}
\author{A.M. Rogers}
\affiliation{National Superconducting Cyclotron Laboratory, Michigan State University, East Lansing, MI 48864, USA}
\affiliation{Department of Physics and Astronomy, Michigan State University, East Lansing, MI 48864, USA}
\author{P. Russotto}
\affiliation{INFN, Sezione di Catania, I-95123 Catania, Italy}
\author{A. Sanetullaev}
\affiliation{National Superconducting Cyclotron Laboratory, Michigan State University, East Lansing, MI 48864, USA}
\affiliation{Department of Physics and Astronomy, Michigan State University, East Lansing, MI 48864, USA}
\author{R. H. Showalter}
\affiliation{National Superconducting Cyclotron Laboratory, Michigan State University, East Lansing, MI 48864, USA}
\affiliation{Department of Physics and Astronomy, Michigan State University, East Lansing, MI 48864, USA}
\author{L.G. Sobotka}
\affiliation{Department of Chemistry, Washington University, St. Louis, MO 63130, USA}
\author{Z.Y. Sun}
\affiliation{National Superconducting Cyclotron Laboratory, Michigan State University, East Lansing, MI 48864, USA}
\affiliation{Institute of Modern Physics, CAS, Lanzhou 730000, Peoples Republic of China}
\author{A.M. Vander Molen}
\affiliation{National Superconducting Cyclotron Laboratory, Michigan State University, East Lansing, MI 48864, USA}
\author{G. Verde}
\affiliation{INFN, Sezione di Catania, I-95123 Catania, Italy}
\author{M.S. Wallace}
\affiliation{National Superconducting Cyclotron Laboratory, Michigan State University, East Lansing, MI 48864, USA}
\affiliation{Department of Physics and Astronomy, Michigan State University, East Lansing, MI 48864, USA}
\author{J. Winkelbauer}
\affiliation{National Superconducting Cyclotron Laboratory, Michigan State University, East Lansing, MI 48864, USA}
\affiliation{Department of Physics and Astronomy, Michigan State University, East Lansing, MI 48864, USA}

\date{\today}

\begin{abstract}
We show that ratios of light-particle energy spectra display scaling properties that can be accurately described by effective local chemical potentials.
This demonstrates the equivalence of $t/^{3}He$ and $n/p$ spectral ratios and provides an essential test of theoretical predictions of isotopically
resolved light-particle spectra.
In addition, this approach allows direct comparisons of many theoretical $n/p$ spectral ratios to experiments where charged-particle spectra
but not neutron spectra are accurately measured. Such experiments may provide much more quantitative constraints on the density and momentum dependence of the symmetry energy.
\end{abstract}

\maketitle

One of the main uncertainties in the Equation  of State (EoS) of neutron-rich nuclear  matter concerns   the density  dependence of the nuclear  symmetry  energy ~\cite{Steiner:2004fi,Danielewicz:2002pu,Tsang:2008fd,Tsang:2012se,Lattimer:2000nx}. The symmetry energy increases  quadratically
with the isospin asymmetry $\delta=(\rho_n-\rho_p)/(\rho_n +\rho_p)$ where $\rho_n$  and  $\rho_p$  are  the neutron  and  proton  densities,  respectively~\cite{Steiner:2004fi,Danielewicz:2002pu,Tsang:2008fd,Tsang:2012se,Lattimer:2000nx,Li:2008gp}.
Stable nuclei therefore have small asymmetries. Neutron stars display large asymmetries $\delta \approx 0.8-0.9$,
driven by the interplay of the Coulomb and symmetry potentials. Accordingly, the density and momentum dependence of the latter strongly influences the internal structure and stability of neutron stars and  their temperatures when formed during core-collapse supernovae~\cite{Lattimer:2000nx,Pons:1998mm}.

Nucleus-nucleus collisions provide  the only  means  to probe  the EoS  of nuclear matter in the laboratory at densities both below \emph{and above}  saturation density.
The symmetry mean-field potential repels neutrons from and attracts protons to a neutron-rich system, leading to predictions that the ratios of neutron (n)  over proton (p)  center of mass  energy  spectra provide  a sensitive probe  of the density and  momentum dependence  of the symmetry energy ~\cite{Rizzo:2005mk,DiToro:2010ku}.

Due to difficulties in precisely knowing neutron detection efficiencies and background corrections, ratios of charged-particle spectra, such as triton (t)
over helion ($^{3}He$) spectra, have been proposed as a more accurate surrogate for n/p spectral ratio measurements. The equivalence of $t/^{3}He$ and n/p spectral ratios,
however, has not been experimentally demonstrated until now.
Progress requires a better understanding of the production of light clusters.

Many theoretical models directly calculate only the ``primordial'' nucleon (n and p) spectra prior to cluster formation and do not make direct predictions
of bound charged-particle ``cluster'' spectra~\cite{Danielewicz:1991dh,Li:2008gp,Chen:2003qj}. Predictions of light ``clusters'', e.g. t or $^{3}He$  spectra,
have been obtained for some models by using the ``coalescence'' approximation of Butler and Pearson~\cite{Butler:1963pp,Danielewicz:1991dh},
or by using other approximations involving the calculated phase-space densities of nucleons~\cite{Chen:2003qj,Aichelin:1991xy,Colonna:2010av}
as they decouple from the system.  The uncertainties of such approximations strongly influence the accuracy of the symmetry energy information that can be extracted
from $t/^{3}He$ spectral ratios. In many cases more reliable conclusions may be drawn by comparing ratios of measured ``coalescence invariant'' n and p spectra
to corresponding ratios calculated by transport models \cite{Danielewicz:2002pu,Famiano:2006rb}. Such comparisons are more analogous to comparisons of "primordial" nucleon spectra and are therefore less sensitive to approximations in the theoretical cluster production mechanisms.
\begin{figure*}[ht]
  \includegraphics[width=.92\textwidth]{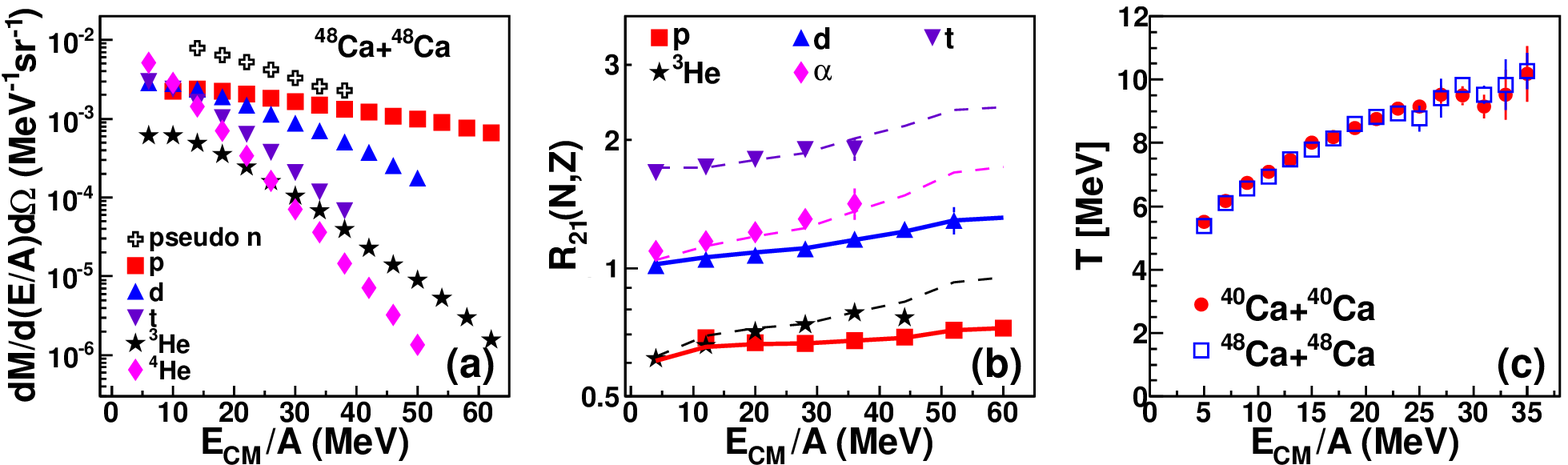}
  \caption{\label{fig:CaFigures}(Color online) (a) Differential multiplicity distributions for \CaFECaFE~collisions at E/A=80MeV, (b) isoscaling yield ratios and (c) the $He-H$ chemical temperature
(Eq.~\ref{eq:Temperature}) from $\ensuremath{^{\text{48,40}}\text{Ca}+^{\text{48,40}}\text{Ca}}$ collisions at $E/A=80$ MeV. The lines in panel (b) represent the fits using Eq.~\ref{eq:fullratio}.
}
\end{figure*}

In this Letter, we clarify the relationships between spectral ratios by showing that all ratios of light-particle center-of-mass energy spectra can be accurately described by two scaling parameters,
related to isoscaling~\cite{Tsang:2001jh},  that depend on the velocities of the particles.
Some theoretical statistical and  dynamical studies relate isoscaling parameters to effective chemical  potentials
~\cite{DiToro:2010ku,Tsang:2001dk,Ono:2003zf}.
The present work shows that assuming effective chemical potentials allows more extensive and more powerful comparisons  between transport theories
and light charged-particle data, facilitating the determination of constraints on the symmetry energy.

We first illustrate these scaling properties by an analysis of experimental measurements of central \CaFECaFE~and \CaFCaF~collisions. In the experiment, isotopically enriched \nuc{48}{Ca} and \nuc{40}{Ca} targets were bombarded with 80 MeV/u \nuc{48}{Ca} and \nuc{40}{Ca} beams, respectively. Hydrogen (p,d,t) and helium ($^{3}He$,$\alpha$)  isotope spectra were measured with the High Resolution Si Strip array  (HiRA)~\cite{hira07}, which subtended angles of $20^{o}\leq\theta_{LAB}\leq60^{o}$, centered about $\theta_{CM}\approx 90^{o}$. The MSU 4$\pi$ array~\cite{Westfall:1985zf} was used to select impact parameters with $b<3 fm$. Further experimental details can be found in Ref.\cite{Henzl:2011dh}.

\begin{figure*}[t]
  \includegraphics[width=.99\textwidth]{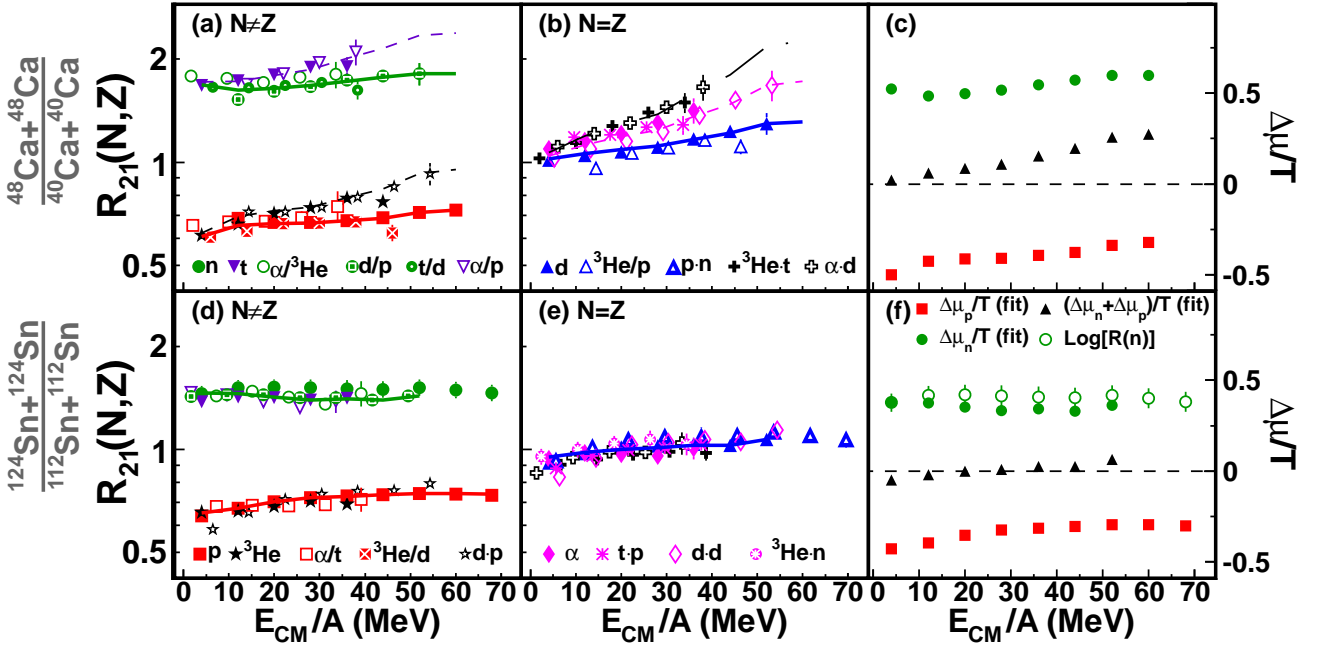}
  \caption{\label{fig:Ratios}(Color online) Isoscaling yield ratios as well as their combinations for (a) and (d) $N \neq Z$ fragments, (b) and (e)  $N=Z$ fragments,
   (c) and (f) overall fit results (solid symbols) using Eq.~\ref{eq:fullratio}. The data points representing the combinations of isoscaling yield ratios in panels (a),(b), (d) and (e) were slightly offset with resect to their average values of $E_{CM}/A$ to improve the readability of the figure.
   The open symbols in (f) represent the values calculated from the measured neutron ratios. The lines in (a),(b),(d) and (e) represent the fit results using Eq.~\ref{eq:fullratio}. The neutron ratio and the combinations of ratios that involve neutrons were not included in the fit, consistent with the treatment for Ca+Ca reactions. The top and bottom panels show results from the $\ensuremath{^{\text{48,40}}\text{Ca}+^{\text{48,40}}\text{Ca}}$ and $\ensuremath{^{\text{124,112}}\text{Sn}+^{\text{124,112}}\text{Sn}}$ collisions, respectively.}
\end{figure*}

Transport calculations predict that the overlap of projectile  and target nuclei in central collisions produces a compressed ``participant'' region whose subsequent expansion reflects the pressure due to the symmetry energy. This pressure can be probed by measurements of energy spectra perpendicular to the beam in the center-of-mass frame, minimizing contributions from the projectile and target residues. The energy spectra for hydrogen and helium isotopes
 for these systems are found to be nearly independent of the center-of-mass scattering angle in the angular range of 70\degree $< \theta_{CM}< $ 110\degree.
The left panel of Fig.~\ref{fig:CaFigures} shows
energy spectra, $\frac{dM(N,Z)}{dE_{CM}d\Omega}$  for \CaFECaFE,  averaged
over this  angular  range  to achieve  higher statistical precision.
All the spectra decrease exponentially with increasing energy. The corresponding spectra for \CaFCaF~display similar trends. These  similarities can be easily observed in the middle panel of Fig.~\ref{fig:CaFigures}, where we show the isoscaling yield ratios
\begin{equation}
\label{eq:fullratio}
R_{21}(N,Z)= \frac{dM_{2}(N,Z)}{dM_{1}(N,Z)}=\mathrm{exp}\Big[N\alpha+Z\beta\Big]
\end{equation}
as a function of $E_{CM}/A$.
In accordance with the isoscaling ratio convention adopted in previous studies~\cite{Tsang:2001dk,Tsang:2001jh}, system 2 represents the more neutron rich reaction, \CaFECaFE, and system 1 represents the neutron deficient reaction, \CaFCaF.
The ratios in the middle panel  separate into three groups: $R_{21}(t) \approx 1.8$ with $N-Z=1$, $R_{21}(d)\approx R_{21}(\alpha)\approx 1.1$ with $N=Z$, and $R_{21}(p)\approx R_{21}(^{3}He)\approx 0.7$ with $N-Z=-1$.  Each ratio increases gradually as a function of energy and this grouping becomes less distinct with increasing $E_{CM}/A$.

In a number of statistical and dynamical models~\cite{DasGupta:1981xx,Tsang:2001dk,Ono:2003zf,Tsang:2001jh,Fai:1987aa,Botvina:2001us}, the isoscaling parameters  $\alpha$  and  $\beta$
can be related to the differences in \emph{effective} neutron and proton chemical potentials, $\mu_n$ and $\mu_p$, between reactions 2 and 1, as well as the effective chemical temperature T.
In such cases, $\alpha = \Delta\mu_n/T$ and $\beta=\Delta\mu_p/T$ and $\Delta\mu_n = (\mu_{n,2}-\mu_{n,1})$,
$\Delta\mu_p = (\mu_{p,2}-\mu_{p,1})$.  These expressions assume that the effective temperature T is the same in both systems 1 and 2.
To examine this assumption, we use the $p, d, ^{3}He$ and $\alpha$ spectra to construct the He-H chemical temperature, which assumes chemical equilibrium at freezeout~\cite{Xi:1996zz,Tsang:1997zz,Albergo:1985zz,Natowitz:2001cq}:
\begin{equation}
\label{eq:Temperature}
T = \frac{14.3}{log[1.59 R_{He-H}]},
\end{equation}
where $R_{He-H}=\frac{dM_{d}/d\Omega dE \cdot dM_{^{4}He}/d\Omega dE}{ dM_{t}/d\Omega dE \cdot dM_{^{3}He}/d\Omega dE}$.
The sensitivity of the $He-H$ thermometer to T stems from the large, $21.6$ MeV, binding energy difference between $\alpha$ and $^{3}He$ clusters, which favors $\alpha$
relative to $^{3}He$ production at low temperature.
The right panel of Fig.~\ref{fig:CaFigures} shows that the extracted He-H isotope temperature values
at $\theta_{CM} \approx 90^{o}$ for \CaFCaF~collisions (full red circles) and \CaFECaFE~collisions (open blue squares) are indistinguishable,
which can be expected for equal center-of-mass kinetic energies per nucleon for these two systems~\cite{Kunde:1998ab}. The monotonic decrease of T with decreasing $E_{CM}/A$ reflects the cooling of the participant source at later stages of the expansion.

The divergent trends of each isotope in the middle panel of Fig.~\ref{fig:CaFigures}  can be described accurately by two isoscaling parameters $\alpha$ and $\beta$,
which both depend on the $E_{CM}/A$ of the outgoing particle.
Accordingly, fits  to $R_{21}(p), R_{21}(t), R_{21}(^{3}He)$ and $R_{21}(\alpha)$ are shown as the lines in  middle  panel  of  Fig.~\ref{fig:CaFigures}.
These  fits agree very well with the data.
This agreement implies fundamental equalities such as  $R_{21}(N_{1}+N_{2},Z_{1}+Z_{2})=R_{21}(N_{1},Z_{1}) \cdot R_{21}(N_{2},Z_{2})$ and $R_{21}(N_{1}-N_{2},Z_{1}-Z_{2})=R_{21}(N_{1},Z_{1})/R_{21}(N_{2},Z_{2})$ between $R_{21}(N,Z)$ values for the various isotopes. For example, it predicts that
$R_{21}(n)=R_{21}(d)/R_{21}(p)=R_{21}(t)/R_{21}(d)$, which is a breakthrough that could obviate part of the need to measure neutrons,
and facilitate more quantitative comparisons the between measured and calculated charged particle spectra.

From the measured spectra of p,d,t,$^{3}He$ and alpha particles, we can test the consistency of n-like ratios  with N-Z=1: $R_{21}(n)$,
$R_{21}(t)$,$R_{21}(\alpha)/R_{21}(^{3}He)$, $R_{21}(d)/R_{21}(p)$, $R_{21}(t)/R_{21}(d)$, $R_{21}(\alpha)/R_{21}(p)$, and p-like ratios with N-Z=-1: $R_{21}(p)$,
$R_{21}(^{3}He)$, $R_{21}(^{3}He)/R_{21}(d)$, $R_{21}(d)R_{21}(p)$ and $R_{21}(\alpha)/R_{21}(t)$
which are shown in the left top panels of Fig.~\ref{fig:Ratios}.
We also construct the deuteron-like ratios with N=Z: $R_{21}(d)$, $R_{21}(\alpha)$, $R_{21}(p)R_{21}(n)$, $R_{21}(^{3}He)/R_{21}(p)$, $R_{21}(^{3}He)R_{21}(t)$, $R_{21}(\alpha)R_{21}(d)$, $R_{21}(t)R_{21}(p)$, $R_{21}(d)R_{21}(d)$, $R_{21}(^{3}He)R_{21}(n)$ as shown in Fig.~\ref{fig:Ratios} (b).  The  isoscaling parameters $\alpha=\Delta\mu_n/T$ and $\beta=\Delta\mu_p/T$ shown in Fig.~\ref{fig:Ratios}(c) were  determined  from  the fits  to all  the distributions  presented  in Fig.~\ref{fig:Ratios}(a) and (b) using Eq.~\ref{eq:fullratio}.

\begin{figure*}[ht]
  \includegraphics[width=.92\textwidth]{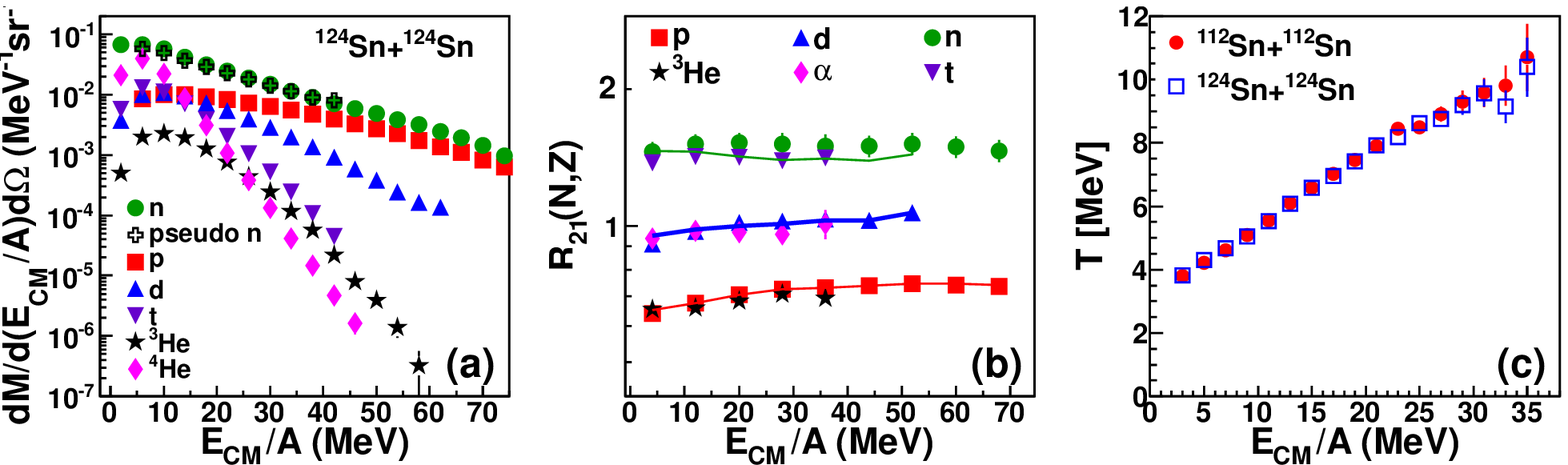}
  \caption{\label{fig:SnFigures}(Color online) (a) Differential multiplicity distributions for \SnSnF~collisions at E/A=50MeV, (b) isoscaling yield ratios and (c) the $He-H$ chemical temperature
(Eq.~\ref{eq:Temperature}) from $\ensuremath{^{\text{124,112}}\text{Sn}+^{\text{124,112}}\text{Sn}}$ collisions at $E/A=50$ MeV. The lines in panel (b) represent the fits using Eq.~\ref{eq:fullratio}.
}
\end{figure*}

To test the applicability of such analyses to predict energy spectra of
neutrons, which were not measured in Ca+Ca experiment, we next investigate n, p, d, t, $^{3}He$ and $^{4}He$ energy spectra
measured for central \SnSnT~and \SnSnF~collisions at E/A=50 MeV from another experiment at
the Coupled Cyclotron Facility (CCF) at Michigan State University~\cite{Coupland2013:InPrep}. In this experiment, we selected central collisions, $b<3 fm$ by measuring the charged-particle multiplicity  and transverse energy using the MSU Miniball Array\cite{DeSouza:1990ry}.
Spectra for hydrogen and helium isotopes were measured with 6 telescopes of the Large Area Silicon Strip Array (LASSA)~\cite{Davin2001302}
and neutrons were measured with the Large Area Neutron Array (LANA)~\cite{Zecher1997329}. Further experimental details can be found in Ref.~\cite{Coupland2013:InPrep}.

The left panel of Fig.~\ref{fig:SnFigures} shows the energy spectra, $\frac{dM(N,Z)}{dE_{CM}d\Omega}$, measured for \SnSnF~collisions, and
the right  panel  of  Fig.~\ref{fig:SnFigures} shows the corresponding  equivalence of temperature values  for  \SnSnT~collisions  (full  red   circles)   and
\SnSnF~collisions  (open  blue  squares). 
The middle panel of Fig.~\ref{fig:SnFigures} shows the ratios $R_{21}(N,Z)$ (Eq.~\ref{eq:fullratio}) of the energy spectra for \SnSnF~collisions divided by those for \SnSnT~collisions.
Similar to Ca+Ca systems, the Sn+Sn ratios vary gradually with energy and separate into three groups: $R_{21}(n) \approx R_{21}(t) \approx 1.5$ with $N-Z=1$, $R_{21}(d)\approx R_{21}(\alpha)\approx 1$ with $N=Z$, and $R_{21}(p)\approx R_{21}(^{3}He)\approx 0.7$ with $N-Z=-1$.
Unlike Ca+Ca, however, $R_{21}(d) \approx R_{21}(\alpha)$ and the various ratios constructed from the measured fragments for
Sn+Sn follow the trends of $R_{21}(n)$, $R_{21}(d)$ and $R_{21}(p)$ more closely; see Figs.~\ref{fig:Ratios}(d) and (e).
Consistent with this observation,
the fit parameters $\alpha=\Delta\mu_n/T$ and $\beta=\Delta\mu_p/T$ in Fig.~\ref{fig:Ratios}(f) satisfy $(\alpha+\beta)\approx 0$ for these Sn+Sn collisions.
The identities $R_{21}(p)=R_{21}(^{4}He)/R_{21}(t)=R_{21}(^{3}He)/R_{21}(d)$ and $R_{21}(n)= R_{21}(d)/R_{21}(p)=R_{21}(t)/R_{21}(d)$ apply to Sn+Sn collisions in Fig.~\ref{fig:Ratios} (d)-(e) as well as they do to Ca+Ca collisions in Fig.~\ref{fig:Ratios}(a)-(b). These scaling behaviors indicate that light-particle emission can be well described by local chemical equilibrium.

To successfully extract symmetry energy constraints directly from isotopically resolved energy spectra, theories should replicate these scaling properties and reproduce the abundances of isoscalar $\alpha$ and d particles at the $E_{CM}/A$ values of interest, since the yields of such isoscalar particles correlate positively  with the n/p and $t/^{3}He$ ratios due to the constraints of charge and baryon number conservation~\cite{Shi:2000ar,Sobotka:1997zz}. If theories fail these tests, comparisons can nevertheless be made between theoretical and experimental ``coalescence invariant'' primordial neutron and proton spectra constructed by combining free nucleons with those bound in clusters. This requires both isotopically resolved charged-particle and free neutron spectra, but few experiments  measure neutrons.

To address the lack of neutron data for central heavy ion collisions, one can use chemical potential scaling to provide relatively accurate estimates for neutron spectra. Applying chemical potential scaling to ratios of spectra within the same isospin multiplet and in the same reaction, chemical potential scaling predicts that
the product  of the measured  triton over helion spectra times the measured  proton spectra provides a ``pseudo''  neutron  spectrum, i.e. $ \frac{dM(n)_{pseudo}}{dE}=\frac{dM(p)}{dE}\cdot \frac{dM(2,1)}{dM(1,2)}$. A somewhat similar procedure has also been proposed in Ref.~\cite{Hagel:2000gj}.
The open black crosses in Fig.~\ref{fig:CaFigures}(a) and~\ref{fig:SnFigures}(a) show ``pseudo'' neutron spectra for the \CaFECaFE~and \SnSnF~reaction respectively.
We compare ``pseudo'' (open black crosses) and free neutron (green circles) spectra in Fig.~\ref{fig:SnFigures}(a);
these two spectra agree to within $10\%$ for  $E_{CM}/A > 20$ MeV. Larger discrepancies might be expected at low energies, due to the Coulomb repulsion from the residue and the difference
between the charges for the n, p, t and $^{3}He$ ejectiles. However, significantly large residues do not survive in central Sn+Sn  collisions at E/A=50 MeV, reducing such effects~\cite{Hudan:2002tn,Liu:2006xs}.
Free or pseudo neutron spectra provide the large contributions to the coalescence invariant neutron spectra at $E_{CM}/A>20~\text{MeV}$, but not at lower energies where
more neutrons are directly observed in clusters, reducing uncertainties due to Coulomb effects. For such reactions where neutrons have not been measured, pseudo-neutron spectra can allow more quantitative constraints on the momentum dependence of the symmetry energy to be extracted, accelerating progress in this important area.

 When local chemical equilibrium is achieved, models predict the effective chemical potential scaling parameters to contain relevant information about the symmetry mean field potentials.  For example, statistical expressions for these scaling parameters $\alpha=\Delta\mu_n/T$ and  $\beta=\Delta\mu_p/T$ can be found in Ref.~\cite{Tsang:2001dk}.
In the Expanding Evaporation Source model (EES), for example, $\alpha$ depends on the neutron separation energy, the excitation energy and entropy per neutron, and $\beta$ depends
on the corresponding proton properties~\cite{Tsang:2001dk}. The separation energies clearly depend on the density and momentum dependence of the symmetry energy within the source,
and the entropies and excitation energies are strongly dependent on its momentum dependence. In the environment of a collision like the ones investigated here, transport theory predicts these sensitivities to the symmetry energy to remain,
but the sensitivity to the momentum dependence also reflects the non-thermal relative nucleonic velocities that requires transport theory for this description ~\cite{Rizzo:2005mk,Das:2002fr}.

In summary,  we have shown that the energy spectra of nucleons and light bound nuclei  follow scaling laws related to isoscaling and to local chemical potentials. This provides an important test of transport theory and confirms the equivalence of n/p to $t/^{3}He$ spectral ratios for systems that totally disintegrate reducing the differences between Coulomb barriers for such particles. We discuss the importance of avoiding the limitations of the cluster production mechanisms of certain models by constructing coalescence invariant primordial neutron and proton spectra. Such spectra  are less sensitive to the final state interactions that produce the clusters  observed  in experiment. We have successfully applied chemical potential scaling to individual reactions to accurately predict the neutron spectra. This will expand considerably the systems from which constraints on the symmetry energy can be obtained accelerating progress in this area.

This work is supported by Michigan State University, the Joint Institute for Nuclear Astrophysics, the National Science Foundation Grants No. PHY-0216783, PHY-1102511,
PHY-1068571, PHY-0822648, and PHY-0855013, and the U.S. Department of Energy Division of Nuclear Physics Grant No. DE-FG02-87ER-40316 and Contract No. DE-AC02-06CH11357.

\bibliographystyle{apsrev}

\end{document}